\documentstyle[12pt]{article}
\newcommand{\beq}{\begin{equation}}
\newcommand{\eeq}{\end{equation}}

\def\half{{\textstyle{1\over2}}}

\def\p1half{{\textstyle{{{p+1}\over{2}}}}}

\def\23phalf{{\textstyle{{{23-p}\over{2}}}}}

\begin{document}
\thispagestyle{empty}
\begin{titlepage}

\bigskip
\hskip 3.7in{\vbox{\baselineskip12pt
}}

\bigskip\bigskip\bigskip\bigskip
\centerline{\large\bf Finite Temperature Gases of Fermionic
Strings}

\bigskip\bigskip
\bigskip\bigskip
\centerline{\bf Shyamoli Chaudhuri
\footnote{shyamoli@thphysed.org}
}
\centerline{214 North Allegheny St.}
\centerline{Bellefonte, PA 16823}
\date{\today}

\bigskip\bigskip
\begin{abstract}
We show that in the absence of a Ramond-Ramond sector both
the type IIA and type IIB free string gases have a thermal
instability due to low temperature tachyon modes.
The gas of free IIA strings undergoes a thermal duality
transition into a gas of free IIB strings at the self-dual
temperature.
The free heterotic string gas is a tachyon-free ensemble with
gauge symmetry $SO(16)$$\times$$SO(16)$ in the presence of
a timelike Wilson line background. It
exhibits a holographic duality relation undergoing
a self-dual phase transition with positive free energy
and positive specific heat.
The type IB open and closed string ensemble is related by
 thermal
duality to the type I$^{\prime}$ string ensemble.
We identify the order
parameter for the Kosterlitz-Thouless phase transition from
a low temperature gas of short open strings to a high temperature
long string phase at or below $T_C$. Note Added (Sep 2005).
\end{abstract}

\end{titlepage}

\section{Introduction}

In this paper we will clarify the properties of the thermal ensemble in
free fermionic string theories--- type IIA, type IIB, heterotic, and
type I, previously studied in \cite{aw,poltorus,hagedorn,salom,tan,decon}.
We emphasize that our considerations \cite{decon,bosonic}
focus on {\em free} strings. All of
our calculations will be based on the one-loop amplitude in string theory
and are valid independent of the strength of the string coupling constant.
Our results for finite temperature string theory are therefore expected
to dovetail neatly with any nonperturbative calculations that follow
in the future.

\vskip 0.1in
The statistical mechanics of an ensemble of free one-dimensional objects
with Planck scale string tension is not the same as that of an ensemble of
infinitely many free particle species with Planck scale masses. It is this
fundamental difference that is stressed in our analysis \cite{decon,bosonic}.
Thus, we will discard previous results based on the assumption of a Boltzmann
distribution since such analyses inherently assume a {\em particle}
ensemble.  Instead, we will take as starting point the
generating functional of connected one-loop vacuum string graphs,
$W(\beta)$ $\equiv$ ${\rm ln}$ ${\rm Z}(\beta)$, known to be
unambiguously normalized in string theory \cite{poltorus}. The following
relations hold:
\begin{equation}
F= -W/\beta = V \rho , \quad P =
      - \left ( {{\partial F}\over{\partial V}} \right )_T , \quad
      U = T^2 \left ( {{\partial W}\over{\partial T}} \right )_V
, \quad
      S = - \left ( {{\partial F}\over{\partial T}} \right )_V ,
 \quad
      C_V = T \left ( {{\partial S}\over{\partial T}} \right )_V
\quad .
\label{eq:freene}
\end{equation}
$W(\beta)$ is an intensive thermodynamic variable without explicit
dependence on the spatial volume.
$F$ is the Helmholtz free energy of the ensemble of free strings,
$U$ is the internal energy, and $\rho$ is the finite temperature
effective potential. $S$ and $C_V$ are, respectively, the entropy
and specific heat of the thermal ensemble. The pressure of the
free string gas vanishes identically such that the enthalpy equals
the energy, $H$$=$$U$, and, consequently, the Gibbs and
Helmholtz free energies coincide: $G$$=$$H$$-$$TS$$=$$U$$-TS$$=$$F$.
Notice that, as a result of these relations, all of the
thermodynamic potentials of a gas of free strings have a simple
representation in terms of worldsheets. The properties of this
canonical ensemble are not those of an ensemble of infinitely many
Planck scale particle species.

\vskip 0.1in
Our main interest is therefore in the generating functional of
connected one-loop vacuum string graphs for the type IIA,
type IIB, heterotic, and type I strings. Unlike the pedagogical--- but
also unphysical, case of the bosonic string gas, the fermionic string
theories have no tachyon in the zero temperature spectrum. It would
therefore be unphysical to have a tachyonic instability at infinitesimal
temperature. We will {\em require}
that our result for the vacuum functional in each case
interpolates smoothly between
the known supersymmetric zero temperature limit and a
tachyon-free thermal description at finite temperature.
For the type II string gas in the absence of Ramond
Ramond backgrounds, we will find no modular invariant and
tachyon free
solutions to these criteria except within a temperature interval
in the vicinity of the string scale. For the
heterotic string gas we find a unique solution with gauge symmetry
$SO(16)$$\times$$SO(16)$. The free energy and vacuum cosmological
are nonvanishing at finite temperature. Interestingly, it becomes
evident in the heterotic case that the topology of the Euclidean time direction is
necessarily that of a particular ${\rm Z}_2$ orbifold. Since this
feature of our analysis is likely to come to
the reader as a surprise, we repeat some of our motivational
statements from a previous analysis of the bosonic string gas
below. Finally, for the open and closed type I string gas, we
find a tadpole and tachyon free solution with $O(16)$$\times$$O(16)$
gauge symmetry.

\vskip 0.1in
Let us note some peculiarities of the thermal prescription for
the free string gases. Since strings are sensitive to the topology
of the underlying target space, the Euclidean time prescription
for the thermal ensemble appears at first sight to be ambiguous: a
one-dimensional compact space can have the topology of a
circle or that of an interval. Which assumption yields the
correct answer for the free energy of a gas of free
strings? The interval implies modding by the ${\rm Z}_2$ twist
in Euclidean time which
breaks \lq\lq time reversal invariance". This symmetry has
no meaningful correspondence in the thermal prescription and
we will therefore eliminate it.
In the closed fermionic strings--- IIA, IIB, or heterotic, the
${\rm Z}_2$ orbifold twist in Euclidean time naturally
extends to a corresponding ${\rm Z}_2$ superorbifold twist, identifying
an appropriate line of ${\hat c}$ $=$ $1$ superconformal
field theories parameterized by interval length, or inverse
temperature: $\beta$ $=$ $\pi r_{\rm circ}$. In the
type I open and closed string gas, the ${\rm Z}_2$ twist acts
as an orientation reversal transformation.

\vskip 0.1in
The physical consequence of modding by a $Z_2$ symmetry
is as follows: a Kosterlitz-Thouless continuous phase transition
occurs at the Hagedorn temperature.
The universality class of the phase transition is indicated
by the existence of analytic expressions for the free energy
and all of its partial derivatives with respect to temperature
upto arbitrary order. This holds for each of the free string
gases.
The physical manifestation of the transition, however, differs
in each case.  In the type II string, the transition maps IIA to
IIB interchanging IIA winding with IIB momentum modes, and vice
versa. In the heterotic string gas, the K-T transition is a
self-duality phase transition leaving the thermal spectrum
invariant. Finally, in the free type I string gas, the
Kosterlitz-Thouless transition marks the transition from a
low temperature phase of short open strings to a high temperature
long string phase. The intuition that a gas of open strings
transitions into a high temperature long string phase is an
old piece of string folklore \cite{polchinskibook,sussk}, with
limited evidence from preliminary analyses of the microcanonical
ensemble \cite{salom}. We will compute the
order parameter for this transition by isolating the
high temperature asymptotics of the pair correlator
of closed timelike Wilson loops. We find a transition
temperature at, or below, $T_C$, in the presence of an
external electric field coupling to open strings alone.
We pause to remark that
this continuous phase transition is the only possible phase
transition compatible with the analyticity of string theory and
consequently, accessible within the worldsheet formalism of
perturbative string theory.

\vskip 0.1in
In summary, while the zero
temperature free string Fock space is supersymmetric and tachyon-free,
an equilibrium thermal ensemble of either closed superstring,
type IIA or type IIB, contains a tachyonic physical state at infinitesimal
temperature \cite{aw}. It follows that, {\em in the absence of gauge
fields, flat spacetime is an unstable ground state for the
ten-dimensional weakly coupled type II string theories under a small
variation in the background temperature.} This result, first pointed
out by us in \cite{decon}, can be stated within the framework of the
renormalization conditions for a consistent finite
temperature closed string perturbation theory: there are no non-tachyonic
solutions to the renormalization conditions in the vicinity
of the zero temperature ground state, in the absence of gauge
fields that can potentially arise in a non-trivial Ramond-Ramond sector.

\vskip 0.1in
On the other hand, in the
presence of a temperature dependent Wilson line, both the 10D
heterotic and type I string theories have {\em a tachyon--free
finite temperature ground state at all temperatures starting from
zero with gauge group $SO(16)$$\times$$SO(16)$.}
The low energy effective theory is a finite temperature gauge-gravity theory,
formulated in modified axial gauge with the Euclidean time component of the
vector potential, $A_0$, set equal to a temperature dependent constant. The
additional massless vector bosons that lead to an enhancement of the
gauge group in the supersymmetric zero temperature limit to either
$E_8$$\times$$E_8$, or $SO(32)$, as appropriate.
We should note here the correspondence of our heterotic Euclidean time
ground state to the Lorentzian signature, nonsupersymmetric and
tachyon-free, 10D heterotic string ground state with
$O(16)$$\times$$O(16)$ found in \cite{sw,agmv,dh,klt} with {\em
positive} cosmological
constant or vacuum energy density.
Arguments similar to those above can explain the zeroes in the
one-loop effective potential of the nonsupersymmetric heterotic
string ground states found in the numerical computations
of \cite{ginine,gv} \cite{itoyama}.

\vskip 0.1in
The plan of this paper is as follows. In section 2.1,
we begin with the adaptation of the GSO conditions to the
Euclidean time prescription for the fermionic string gas.
This is followed in section
2.2 by a derivation of the worldsheet representation
of the oneloop string vacuum functional for an equilibrium ensemble
of either free type IIA or type IIB strings at finite temperature.
In Section 2.3, we address the low temperature thermal
instability of either type II ensemble and derive the upper and
lower bounds on the temperature axis within which the
thermal spectrum is tachyon-free.
A brief discussion of a plausible nonperturbative resolution
of the thermal instability in the presence of Ramond-Ramond
backgrounds ends section 2.

\vskip 0.1in
Section 3 begins with a derivation of the vacuum functional
for the free heterotic string gas. We show that in the presence
of a temperature-dependent Wilson line background, the
thermal spectrum is tachyon-free at all temperatures starting
from zero and has gauge symmetry
$SO(16)$$\times$$SO(16)$.
In section 3.2, we derive the thermodynamic potentials of the
free heterotic string gas demonstrating the holographic
duality relation and the Kosterlitz-Thouless self-duality
transition. We clarify that the entropy is finite with no
divegences, and with no remnant of the unphysical fixed point
entropy found in the pedagogically similar bosonic string
gas \cite{bosonic}.

\vskip 0.1in
Section 4 describes the type I open and closed string gas.
In section 4.1, we obtain the generating functional of oneloop
vacuum string graphs and the free energy. The free energy vanishes as a
consequence of dilaton tadpole cancellation. The order parameter
for the transition to the high temperature long string
phase is derived in section 4.2. In section 4.3, we
evaluate the scaling behavior for the first few thermodynamic
potentials, verifying that the phase transition is in the
universality class of the Kosterlitz-Thouless transition.
The conclusions and a brief discussion of open questions
appear in section 5.

\section{The Free Type II String Gas}

\subsection{Adapting the GSO Projection}

Recall that in the imaginary time formalism, spacetime fields obeying
Fermi statistics are to be expanded in a Fourier basis of antiperiodic
modes alone. This implies antiperiodic boundary conditions for the
spacetime fermionic modes of the string path integral in the imaginary
time direction. The spacetime spin and statistics of fields
in the Euclidean embedding space can be determined in string theory by
a suitable adaptation of the GSO projection, thus determining
the {\em world-sheet} spin and statistics of states in the free string
thermal spectrum. We will use the Ramond-Neveu-Schwarz (RNS) formulation
to compute the thermal spectrum of the fermionic free string gas.
We recall that fermions on the worldsheet are simultaneously spacetime
vectors and worldsheet spinors.

\vskip 0.1in
There is no spin-statistics theorem in two dimensions and we are
therefore free to choose arbitrary boundary conditions on the
world-sheet fermions \cite{sw,ckt}. We emphasize that such a generalized
sum over worldsheet spin structures is {\em necessitated} by the
simultaneous requirements of the worldsheet symmetries--- modular
invariance and superconformal invariance, and the correct spacetime
physics: a non-tachyonic thermal spectrum, with a supersymmetric zero
temperature limit. In practise, rather than spell out the sum over
spin structures and the modified GSO projection,
it is easier to obtain the one-loop amplitude by
manipulating modular invariant blocks of Jacobi theta functions which
preserve spacetime Lorentz invariance, and consequently, the world-sheet
superconformal invariances. Thus, we identify modular invariant
blocks of eight spin structures pertaining
to the eight transverse world-sheet fermions. These blocks are weighted
by, a priori, undetermined temperature-dependent phases.

\vskip 0.1in
Let us move on to a discussion of the type II one-loop vacuum functional.
In the appendix we recall some relevant facts on the possible action of the
superorbifold group, and on modular invariant lattice partition functions.
We will find that the vacuum functional for either IIA or IIB free string
ensemble takes identical form in the absence of RR backgrounds
and Yang-Mills fields. The thermal duality transformation
mapping the IIA string to the IIB string will simply act as an invariance
on this function, mapping IIA winding to IIB momentum modes and
vice versa.
We begin with the expression for the
normalized generating functional of connected one-loop vacuum
graphs:
\begin{equation}
W_{\rm II} =
\int_{\cal F} {{d^2 \tau}\over{4\tau_2^2}}
  (2 \pi \tau_2 )^{-4} |\eta(\tau)|^{-14} Z_{\rm II} (\beta)
\quad ,
\label{eq:typeII}
\end{equation}
where the spatial volume, $V$ $=$ $ L^{9} (2\pi \alpha^{\prime})^{9/2} $,
and the inverse temperature is given by $\beta $ $=$ $ \pi r_{\rm circ.}$.
The function ${\rm Z}_{\rm II~ orb.}(\beta)$ is the product of
contributions from worldsheet fermions and bosons,
${\rm Z}_F Z_{\rm B}$, and is required to smoothly interpolate
between finite temperature and the spacetime supersymmetric
zero temperature limit.
The spectrum of thermal modes is unambiguously determined by modular
invariance.

\vskip 0.1in
Following the appendix, we consider the action of the super-orbifold
group $R (-1)^{N_F}$,
with an accompanying half-momentum or half-winding shift,
$\delta$$=$$\half({{1}\over{x}} +{{x}\over{2}},{{1}\over{x}}-{{x}\over{2}})$,
in the momentum lattice. We have introduced the
dimensionless inverse temperature
(radius) defining $x$ $\equiv$ $r(2/\alpha^{\prime})^{1/2}$,
with $\beta$ $=$ $\pi (\alpha^{\prime}/2)^{1/2} x$. The
dimensionless quantized momenta live in
a $(1,1)$ dimensional Lorentzian self-dual lattice
\cite{nsw,dgh,polchinskibook}:
\begin{equation}
\Gamma_{\delta} (x) \equiv \sum_{w \in {\rm Z}, n \in {\rm Z} }
 q^{{{1}\over{2}} ({{n }\over{x}} + {{wx}\over{2}} )^2 }
 {\bar q}^{{{1}\over{2}} ({{n}\over{x}} - {{wx}\over{2}} )^2 }
 \quad ,
\label{eq:evenoddshift}
\end{equation}
The purpose of the shift is to
introduce a positive, temperature-dependent,
shift in the spectrum of thermal masses.
The spacetime supersymmetry breaking orbifold action $(-1)^{N_F}$ is
modified by the introduction of {\em phases} in the interpolating
function. Such phases can depend on thermal mode number. They must
be chosen {\em compatible with the requirement
that spacetime supersymmetry is restored in the zero temperature limit
of the interpolating function}.
We comment that temperature dependent phases were first proposed
in \cite{aw}.

\vskip 0.1in
The {\em unique} modular invariant interpolating function
satisfying these requirements is:
\begin{eqnarray}
Z_{\rm II} =&& {{1}\over{2}} {{1}\over{\eta{\bar{\eta}}}} \left [
\sum_{w,n \in {\rm Z} }
   q^{{{1}\over{2}} ({{n }\over{x}} + {{wx}\over{2}} )^2 }
     {\bar q}^{{{1}\over{2}} ({{n}\over{x}} - {{wx}\over{2}} )^2 }
 \right ]
\{ (|\Theta_3 |^8
  + |\Theta_4|^8  + |\Theta_2|^8 ) \cr
  \quad&&\quad\quad + e^{ \pi i (n + w ) }
 [ (\Theta_2^4 {\bar{\Theta}}^4_4 + \Theta_4^4 {\bar{\Theta}}_2^4)
  - (\Theta_3^4 {\bar{\Theta}}_4^4 + \Theta_4^4 {\bar{\Theta}}_3^4
   + \Theta_3^4 {\bar{\Theta}}_2^4 + \Theta_2^4 {\bar{\Theta}}_3^4 ) ]
\quad .
\label{eq:bosod}
\end{eqnarray}
The world-sheet fermions have been conveniently
complexified into left- and right-moving
Weyl fermions. As in the superstring, the spin structures for all ten left- and
right-moving fermions, $\psi^{\mu}$, ${\bar{\psi}}^{\mu}$, $\mu$
$=$ $0$, $\cdots$, $9$, are determined by those for the world-sheet gravitino
associated with left- and right-moving N=1 superconformal invariances.

\vskip 0.1in
To understand our result for the correct interpolating function, first recall
the expression for ${\rm Z}_{SS}$--- the zero temperature, spacetime
supersymmetric, limit of our function given by the ordinary GSO projection:
\begin{equation}
 Z_{\rm SS} =
     {{1}\over{4}}  {{1}\over{\eta^4{\bar{\eta}}^4}}
  \left [ (\Theta_3^4 - \Theta_4^4 - \Theta_2^4)
   ({\bar{\Theta}}_3^4 - {\bar{\Theta}}_4^4 - {\bar{\Theta}}_2^4 )
\right ]
 \quad ,
\label{eq:IIs}
\end{equation}
Notice that the first of the relative signs
in each round bracket preserves the tachyon-free condition. The second relative sign
determines whether spacetime supersymmetry is preserved in the zero
temperature spectrum. Next, notice that $Z_{SS}$ can be rewritten
using theta function identities as follows:
\begin{eqnarray}
 Z_{\rm SS } &&=
\{ [ |\Theta_3 |^8
  + |\Theta_4|^8  + |\Theta_2|^8 ] \cr
  \quad&&\quad\quad +
 [ (\Theta_2^4 {\bar{\Theta}}^4_4 + \Theta_4^4 {\bar{\Theta}}_2^4)
  - (\Theta_3^4 {\bar{\Theta}}_4^4 + \Theta_4^4 {\bar{\Theta}}_3^4
   + \Theta_3^4 {\bar{\Theta}}_2^4 + \Theta_2^4 {\bar{\Theta}}_3^4 ) ]
  \}
 \quad .
\label{eq:IIid}
\end{eqnarray}
Either of the two expressions within square
brackets is modular invariant.
The first may be recognized as the nonsupersymmetric sum over
spin structures for the type 0 string \cite{polchinskibook}.

\vskip 0.1in
Thus, the interpolating function captures the desired zero
temperature limits of both the IIA and IIB strings: for
large $\beta_{IIA}$, terms with $w_{IIA}$$\neq$$0$
decouple in the double summation since they are exponentially
damped. The remaining terms are resummed by a Poisson
resummation, thereby inverting the $\beta_{IIA}$ dependence in the
exponent, and absorbing the phase factor in a shift of argument.
The large $\beta_{IIA}$ limit can then be taken smoothly
providing the usual momentum integration for a non-compact dimension,
plus a factor of $\tau_2^{-1/2}$ from the Poisson resummation.
This recovers the supersymmetric zero temperature partition function.

\vskip 0.1in
As regards thermal duality, small $\beta_{IIA}$ maps to large
$\beta_{IIB}$$=$$\beta_C^2/\beta_{IIA}$, also interchanging the
identification of momentum and winding modes,
$(n,w)_{IIA}$$\to$$(n'=w,w'=n)_{IIB}$.
We can analyze the limit of large dual
inverse temperature as before, obtaining the zero temperature
limit of the dual IIB theory.
At any intermediate temperature, all of the
thermal modes contribute to the vacuum functional with
a phase that takes values $(\pm 1)$ only. Note that the spacetime
fermions of the zero temperature spectrum now contribute with
a reversed phase, evident in the first term in
Eq.\ (\ref{eq:bosod}), as required by the
thermal boundary conditions.
In summary, the vacuum functional of the IIA string
maps precisely into the vacuum functional of the IIB string
under a thermal duality transformation.

\subsection{Tachyonic Thermal Momentum and Thermal Winding Modes}

\vskip 0.1in
We now point out a peculiarity of the type IIA and IIB free string
gas at finite temperature.
First, recall some pertinent facts from quantum field
theory. It is generally assumed to be the case \cite{kirzhlinde}
that in a field theory known to be perturbatively renormalizable
at zero temperature, the new infrared divergences introduced by a small
variation in the background temperature can be self-consistently
regulated by a suitable
extension of the renormalization conditions, at the cost of introducing a
finite number of additional counter-terms. The zero temperature
renormalization conditions on 1PI Greens functions are conveniently applied
at zero momentum, or at fixed spacelike momentum in the case of massless
fields. Consider a theory with one or more scalar fields. Then the
renormalization conditions
must be supplemented by stability constraints on
the finite temperature effective potential \cite{kirzhlinde}:
\begin{equation}
{{\partial V_{\rm eff.}(\phi_{\rm cl.})}\over{\partial
\phi_{\rm cl.}}} = 0, \quad
\quad
{{\partial^2 V_{\rm eff.}(\phi_{\rm cl. })}\over{\partial \phi_{\rm cl.}^2 }} \equiv
G_{\rm \phi}^{-1} (k)|_{k=0}
\geq 0 \quad .
\label{eq:stab}
\end{equation}
Here, $V_{\rm eff.}(\phi_{\rm cl.})$$=$$-\Gamma(\phi_{\rm cl.})/V\beta$, where
$\Gamma$ is the effective action functional, or sum of
connected 1PI vacuum diagrams at finite temperature. In the absence of
nonlinear field configurations, and for perturbation theory in a small coupling
about the free field vacuum, $|0>$, $\phi_{\rm cl.}$ is simply the
expectation value of the scalar field: $\phi_{\rm cl.}$$=$$<0 | \phi(x) | 0 >$.
The first condition holds in the absence of an external source at every
extremum of the effective potential.
The second condition states that the renormalized masses of physical fields
must not be driven to imaginary values at any non-pathological and stable
minimum of the effective potential.

\vskip 0.1in
The conditions in Eq.\ (\ref{eq:stab}) are, in fact, rather
familiar to string theorists.
Weakly coupled superstring theories at zero temperature are replete with
scalar fields and their vacuum expectation values, or moduli, parameterize a
multi-dimensional space of degenerate vacua. Consider the effect of
an infinitesimal variation in the background temperature. Such an effect will
necessarily break supersymmetry, and it is well-known that
in the presence of a small spontaneous breaking of supersymmetry the dilaton
potential will generically develop a runaway direction, signalling an instability
of precisely the kind forbidden by the conditions that must be met by a
non-pathological ground state \cite{dinesei}. Namely, while the zero temperature
effective potential is correctly minimized with respect to the renormalizable
couplings in the potential and, in fact, vanishes in a spacetime supersymmetric
ground state, one or more of the scalar masses
is driven to imaginary values in the
presence of an {\em infinitesimal} variation in background
temperature. Such a quantum
field theory is simply unacceptable both as a self-consistent effective field
theory in the Wilsonian sense, and
also as a phenomenological model for a
physical system \cite{kirzhlinde}. The same conclusion must hold for a weakly
coupled superstring theory with these pathological properties.

\vskip 0.1in
In the absence of gauge fields, the low temperature behavior of the
ten-dimensional type II superstrings is pathological in the sense described
above \cite{aw,decon}. To check for potential tachyonic instabilities in the
expressions given in Eq.\ (\ref{eq:bosod}), consider the mass
formula in the (NS,NS) sector for world-sheet fermions,
with ${\bf l}_L^2 $ $=$ $ {\bf l}_R^2$,
and $N$$=$${\bar{N}}$$=$$0$:
\begin{equation}
({\rm mass})^2_{nw} =  {{2}\over{\alpha^{\prime}}}
 \left [ - 1 +
{{\alpha^{\prime} \pi^2n^2 }\over{2 \beta^2}} +
{{\beta^2w^2}\over{2 \pi^2 \alpha^{\prime} }}  \right ] \quad .
\label{eq:massII}
\end{equation}
This is the only sector that contributes tachyons to the thermal
spectrum. A nice check is that the momentum mode number
dependence contained in the phase factors
does not impact the spacetime spin-statistics relation in
the NS-NS sector: all potentially
tachyonic states are spacetime scalars as expected.
Notice that the $n$$=$$w$$=$$0$ sector common to both type II
string theories contains a potentially tachyonic state whose
mass is now temperature dependent. However, it corresponds to
an unphysical tachyon. The potentially tachyonic
physical states are the pure momentum and pure winding states,
$(n,0)$ and $(0,w)$, with $N$$=$${\bar{N}}$$=$$0$. The R-orbifold
projects to the symmetric linear combinations of net zero
momentum and net zero winding number states as explained in
\cite{bosonic}.

\vskip 0.1in There is a brief window on the temperature axis for
which the physical state spectrum is tachyon-free. In analogy with
the bosonic string analysis, we can compute the temperatures
below, and beyond, which these modes become tachyonic in the
absence of oscillator excitations. Each momentum mode, $(\pm n,
0)$, is tachyonic {\em upto} some critical temperature, $T^2_n$
$=$ $1/2 n^2\pi^2 \alpha^{\prime} $, after which it becomes
stable. Conversely, each winding mode $(0,\pm w)$, is tachyonic
{\em beyond} some critical temperature, $T^2_w$ $=$ $2w^2/\pi^2
\alpha^{\prime}$. It is evident that the $(1,0)$ and $(0,1)$ modes
determine the upper and lower critical temperatures for a
tachyon-free physical state spectrum. We refer to these,
respectively, as the minimum and maximum temperatures for the
given type II string. A thermal duality transformation maps IIA to
IIB, mapping $T_{\rm min}(IIA)$ to $T_{\rm max}(IIB)$, and vice
versa, since the identification of momentum and winding modes is
switched in the mapping. Expressed in terms of the dual
temperature variable, the minimum and maximum temperatures will,
of course, coincide. As a consequence, the IIA and IIB thermal
strings have common minimum and maximum temperatures. The thermal
mass spectrum of either type II string is tachyon-free for
temperatures within the interval:
\begin{equation}
T_{\rm min} < T < T_{\rm max} , \quad {\rm where}
 \quad T^2_{\rm min} = 1/2\pi^2 \alpha^{\prime },
\quad T^2_{\rm max} = 2/\pi^2 \alpha^{\prime }
 \quad .
\label{eq:rant}
\end{equation}
Notice that there are no additional massless thermal modes at the
self-dual temperature, $\beta_C^2$$=$$\pi^2 \alpha'$, from which
we can infer
that only a one-parameter line of world-sheet superconformal
field theories passes through the self-dual point, parameterized by the
inverse temperature.

\vskip 0.1in
Finally, we remark that the source of the instability discussed here
should not be confused with the gravitational instability of flat spacetime
found in the finite temperature field theory analysis of \cite{gyp}.
Nor is it to be confused with the famous Jeans instability of
gravitating matter in the limit of infinite spatial volume \cite{aw}.
Our considerations have been limited to the internal consistency
conditions of the {\em free} closed string thermal spectrum. In
particular, the ensemble of closed strings does not gravitate at
this order in perturbation theory.

\subsection{Resolution of the Type II Thermal Instability}

\vskip 0.1in
We now point out a plausible resolution of the thermal instability
in the type II string gas. A detailed demonstration of the
validity of our hypothesis takes us outside the realm of
string perturbation theory. While we can infer such a resolution
using our knowledge of the strong-weak effective dualities
that characterize the low energy limits of the type II strings,
we lack at the current time a fully nonperturbative framework
where our proposal can be tested. Fortunately, the mechanism we
propose has a precise analog in the case of the free heterotic
string gas where it can be analyzed in entirety within a
perturbative treatment, and without recourse to conjectural
weak-strong
duality in the absence of supersymmetry. That straightforward
discussion appears in the next section. Let us however summarize
our limited understanding of the type II case below.

\vskip 0.1in
Recall that the low energy limits of the
heterotic and type II string
theories are related by a weak-strong duality in dimensions six and
below, such that Ramond-Ramond ($C_{\mu}$) charge is
identified with a one-dimensional (abelian) subspace in the
$(20,4)$ heterotic momentum lattice \cite{polchinskibook,allCL}.
This low energy weak-strong duality can be lifted up to an
heterotic-M theory duality in dimensions nine and below.
In adapting this to the Euclidean time prescription for both
theories, we
must keep in mind that we have little, or no, control in
making reliable statements about the massive modes of the
string spectrum which may be strongly coupled. However,
our interest here lies in the tachyonic thermal modes, and
the possibility of observing their transition to
masslessness.

\vskip 0.1in
Observe that a timelike Wilson line background for the
Abelian one-form potential of the form:
$C_{[1]}$$=$${{1}\over{\beta}}$, would give rise to a
{\em temperature
dependent shift} in the mass of the thermal tachyonic
modes of the type II string. Such Wilson line backgrounds
for the Ramond-Ramond oneform potential were discussed
in detail in \cite{allCL}. Related work on type 0 orientifolds
\cite{klebt,orient} is also of likely relevance here.
In the dual heterotic string gas, the RR oneform is mapped
to an abelian subspace of the Yang-Mills vector potential.
The role of the timelike Wilson line background for the
gauge field is, likewise, to provide the shift in the
mass of the thermal tachyonic mode. Unlike the type II
case, however, we can analyze the consequences of the
timelike background on the full thermal spectrum including
massive string modes. That is done in the next section.

\vskip 0.1in
We close this discussion
with the comment that the proper context for a
discussion of the type II string gas in the presence of
Ramond-Ramond backgrounds is more precisely
the {\em massive} type II string supergravity:
upon inclusion of a
mass parameter, or cosmological constant, the dimensionally
reduced 10d IIA theory
is equivalent
to the Scherk-Schwarz dimensionally-reduced
massive IIB string supergravity \cite{romans,berg,hull,flux}. The
field content of IIA and IIB theories are identified under the
$T_9$ duality as follows:
\begin{eqnarray}
\{ G_{\mu\nu},\Phi,C_{\mu\nu\lambda},C_{\mu},B_{\mu\nu},
 C_{\mu\nu 9},B_{\mu 9},
{{G_{\mu 9}}\over{G_{99}}},
{\sqrt{-G_{99}}}, C_9 \} \quad\quad&& {\rm Type ~ IIA} \cr
\{ G_{\mu\nu},\Phi, C_{\mu\nu\lambda9},C_{\mu9} , C_{\mu\nu},
B_{\mu\nu},B_{\mu},{\bar {A}}_{\mu},
e^{\chi} , C_{[0]} \} \quad\quad&& {\rm Type ~ IIB} \quad ,
\label{eq:fiecont}
\end{eqnarray}
such that the 9d spacetime actions coincide upto field redefinitions.
From the $SL(2,Z)$
invariance of the IIB theory, we also know that
the mass parameter is quantized
in integer units:
\begin{equation}
C_{[0]} \to C_{[0]} + n , \quad
M = {{ n }\over{R_{9B}}}
\quad , n \in {\rm Z} \quad .
\label{eq:qntcp}
\end{equation}
We note that it is likely that the value of
$n$ is determined by nonperturbative considerations. An argument
based on heterotic-type I duality given by us in \cite{flux}
concludes that the fundamental type II string carries a
single unit, $n$$=$$1$, of Ramond-Ramond scalar charge in the
massive theory. The 9d massive supergravity with its full spectrum
of Dbranes has the remarkable feature that, by field redefinitions
and T-dualities alone, it can be matched with the 9d low energy
limits of any of the supersymmetric string theories--- type
II, heterotic, type I, and type I$^{\prime}$. Further
discussion of the nonperturbative type II string gas must be
deferred to future work.

\section{The Free Heterotic String Gas}

\subsection{The Timelike Wilson Line Background}

\vskip 0.1in
Next consider an equilibrium ensemble of free heterotic
strings in nine noncompact spatial
dimensions occupying the box-regulated volume $V$ $=$
$L^9 (2\pi \alpha^{\prime})^{9/2}$. The normalized
generating functional of connected one-loop
vacuum string graphs is given by the expression:
\begin{equation}
W_{\rm het.} =
\int_{\cal F} {{d^2 \tau}\over{4\tau_2^2}}
  (2 \pi \tau_2 )^{-9/2} |\eta(\tau)|^{-14} Z_{\rm het.} (\beta)
\quad .
\label{eq:het}
\end{equation}
As in the type II case, we wish to identify a modular invariant
interpolating function which matches smoothly with the known,
zero temperature,
spacetime supersymmetric limit:
\begin{equation}
W_{\rm het.}|_{\beta = \infty} =
\int_{\cal F} {{d^2 \tau}\over{4\tau_2^2}}
  (2 \pi \tau_2 )^{-5} |\eta(\tau)|^{-16}
{{1}\over{8}}  {{1}\over{{\bar{\eta}}^4}}
   \{ ({\bar{\Theta}}_3^4 - {\bar{\Theta}}_4^4 - {\bar{\Theta}}_2^4 )
\left [ ({{\Theta_3}\over{\eta}})^8
+ ({{\Theta_4}\over{\eta}})^8
+ ({{\Theta_2}\over{\eta}})^8 \right ]^2 \}
\quad .
\label{eq:holos}
\end{equation}
$Z_{\rm het.}$ describes the mass spectrum of free
$E_8$$\times$$E_8$ strings at finite temperature.
The term enclosed by curly brackets
in the expression above is the
supersymmetric sum over spin structures
given by the chiral GSO projection, $Z_{\rm HSS}$,
including the contribution of the gauge fermions
for the $E_8$$\times$$E_8$ lattice.

\vskip 0.1in
The modular invariant possibilities for the sum over spin structures in
the 10d heterotic string have been classified, both by free fermion and by
orbifold techniques \cite{sw,dh,agmv,klt}, and
there is a {\em unique} tachyon-free solution with
gauge group $SO(16)$$\times$$SO(16)$. It corresponds to an $R(-1)^F$
orbifold of the $E_8$$\times$$E_8$ string. We see that, quite independent
of the motivational statements in our previous analysis of the bosonic and
type II string gases, the thermal formulation of free heterotic string gas
in the Euclidean time prescription {\em requires} that we mod out by
\lq\lq time reversal invariance". Here the ${\rm Z}_2$ action is within
the context of the heterotic superorbifold group. Let us describe how this
works in detail.

\vskip 0.1in
Recall the radius-dependent Wilson line background described by
Ginsparg in \cite{ginine} which provides the smooth interpolation
between the
heterotic $E_8$$\times$$E_8$ and $SO(32)$ theories in nine dimensions.
We have: ${\bf A}$$=$${{2}\over{x}}(1,0^7,-1,0^7)$, $x$$=$$
({{2}\over{\alpha^{\prime}}})^{1/2} r_{\rm circ.}$. Introducing this background
connects smoothly the 9D supersymmetric
$SO(16)$$\times$$SO(16)$ string at generic radii with the
supersymmetric 10d limit where the gauge group is enhanced to
$E_8$$\times$$E_8$. Note that the states in the spinor
lattices of $SO(16)$$\times$$SO(16)$ correspond to
massless vector bosons only in the noncompact limit.
Generically, the $(17,1)$-dimensional heterotic
momentum lattice takes the form
$E_8$$\oplus$$E_8$$\oplus$$U$. Here, $U$ is the
$(1,1)$ momentum lattice corresponding to compactification
on a circle of radius
$r_{\rm circ.}$$=$$x(\alpha^{\prime}/2)^{1/2}$. A generic Wilson
line corresponds to a lattice boost as follows \cite{ginine}:
\begin{equation}
({\bf p}; l_L, l_R) \to
({\bf p}'; l^{\prime}_L, l^{\prime}_R) =
({\bf p} + w x {\bf A};
u_L - {\bf p}\cdot {\bf A}
- {{wx}\over{2}} {\bf A} \cdot {\bf A} ,
u_R - {\bf p}\cdot {\bf A}
- {{wx}\over{2}} {\bf A} \cdot {\bf A} ) \quad .
\label{eq:boost}
\end{equation}
${\bf p}$ is a 16-dimensional lattice vector in
$E_8$$\oplus$$E_8$. As shown in \cite{ginine},
the vacuum functional of the
supersymmetric 9d heterotic string, with generic radius {\em and}
generic Wilson line in the compact spatial direction,
can be written in terms of a sum over
vectors in the boosted lattice:
\begin{equation}
W_{\rm SS} (r_{\rm circ.} ; {\bf A}) =
\int_{\cal F} {{d^2 \tau}\over{4\tau_2^2}}
  (2 \pi \tau_2 )^{-5} |\eta(\tau)|^{-16}
{{1}\over{8}}  {{1}\over{{\bar{\eta}}^4}}
   ({\bar{\Theta}}_3^4 - {\bar{\Theta}}_4^4 - {\bar{\Theta}}_2^4 )
\left [ {{1}\over{\eta^{16}}}
\sum_{({\bf p}'; l^{\prime}_L , l^{\prime}_R)} q^{\half ({\bf p}^{\prime 2} +
  l_L^{\prime 2})} {\bar q}^{\half l_R^{\prime 2}}
\right ]
\quad .
\label{eq:hollat}
\end{equation}
$W_{\rm SS}$ describes the supersymmetric
heterotic string with gauge group $SO(16)$$\times$$SO(16)$ at generic radius.
Now, as explained above, the nonsupersymmetric but tachyon-free 9d string
with gauge group $SO(16)$$\times$$SO(16)$ at generic radii is
given by an $R(-1)^{F}$ orbifold of the compactified supersymmetric
$E_8$$\times$$E_8$ model \cite{dh,itoyama,sw,agmv,klt,gv}):
\begin{equation}
W_{\rm NS} (r_{\rm circ.} = {{1}\over{4}}
\left [ ({{\Theta_2}\over{\eta}})^8
({{\Theta_4}\over{\eta}})^8
({{{\bar{\Theta_3}}}\over{\eta}})^4
- ({{\Theta_2}\over{\eta}})^8
({{\Theta_3}\over{\eta}})^8
({{{\bar{\Theta_4}}}\over{\eta}})^4
- ({{\Theta_3}\over{\eta}})^8
({{\Theta_4}\over{\eta}})^8
({{{\bar{\Theta_2}}}\over{\eta}})^4 \right ]
  \sum_{n,w} q^{\half {\bf l}_L^2 } {\bar{q}}^{\half {\bf l}_R^2}
\quad .
\label{eq:dualh}
\end{equation}
However, by identifying an appropriate interpolating function as
in previous sections and appropriate background field, we can continuously
connect this background to the supersymmetric $E_8$$\times$$E_8$
string.

\vskip 0.1in
We will identify the interval length in Euclidean
time with inverse temperature, such that $x$$=$$
({{2}\over{\alpha^{\prime}}})^{1/2} {{\beta}\over{\pi}}$.
From the viewpoint of the low-energy
finite temperature gauge theory the timelike Wilson line
is simply understood as imposing a modified axial gauge
condition: $A^0$$=$${\rm const}$.
The dependence of the constant on background temperature has been
chosen to provide a shift in the mass formula
that precisely cancels the contribution from low
temperature $(n,0)$ tachyonic modes.
As before, we begin by identifying an appropriate
modular invariant interpolating
function:
\begin{eqnarray}
 Z_{\rm het.} =&& {{1}\over{2}}
\sum_{n,w}
 \left [ ({{\Theta_2}\over{\eta}})^8
({{\Theta_4}\over{\eta}})^8
({{{\bar{\Theta_3}}}\over{\eta}})^4
- ({{\Theta_2}\over{\eta}})^8
({{\Theta_3}\over{\eta}})^8
({{{\bar{\Theta_4}}}\over{\eta}})^4
- ({{\Theta_3}\over{\eta}})^8
({{\Theta_4}\over{\eta}})^8
({{{\bar{\Theta_2}}}\over{\eta}})^4 \right ]
  q^{\half {\bf l}_L^2 } {\bar{q}}^{\half {\bf l}_R^2}
\cr &&   {{1}\over{4}} \sum_{n,w} e^{\pi i (n+w)}\left [ ( {{
{\bar{\Theta_3}} }\over{\eta}} )^4
-({{{\bar{\Theta_2}}}\over{\eta}})^4-({{{\bar{\Theta_4}}}\over{\eta}})^4
\right ] \left [ ({{\Theta_3}\over{\eta}})^{16} + (
{{\Theta_4}\over{\eta}})^{16} + ({{\Theta_2}\over{\eta}})^{16})
\right ] q^{\half {\bf l}_L^2 } {\bar{q}}^{\half {\bf l}_R^2} \cr
\quad && \quad + {{1}\over{2}} \sum_{n,w} e^{\pi i (n+w)}
 \{ ({{{\bar{\Theta_3}}}\over{\eta}})^4 ({{\Theta_3}\over{\eta}})^8
 \left [
({{\Theta_4}\over{\eta}})^8 + ({{\Theta_2}\over{\eta}})^8 \right ]
- ({{{\bar{\Theta_2}}}\over{\eta}})^4 ({{\Theta_2}\over{\eta}})^8
\left [ ({{\Theta_3}\over{\eta}})^8 + ({{\Theta_4}\over{\eta}})^8
\right ] \cr && \quad \quad \quad -
({{{\bar{\Theta_4}}}\over{\eta}})^4 ({{\Theta_4}\over{\eta}})^8
\left [ ({{\Theta_2}\over{\eta}})^8 + ({{\Theta_3}\over{\eta}})^8
\right ] \}
  q^{\half {\bf l}_L^2 } {\bar{q}}^{\half {\bf l}_R^2} \quad .
\label{eq:iden}
\end{eqnarray}
As in previous sections, the first term within square brackets has been
chosen as the
nonsupersymmetric sum over spin structures
for a {\em chiral} type 0 string. This function appears in the
sum over spin structures for the tachyon-free $SO(16)$$\times$$SO(16)$
string given above.
Notice that taking the $x$$\to$$\infty$
limit, by similar manipulations as in the type II case, yields the
partition function of the supersymmetric 10D $E_8$$\times$$E_8$ string.

\vskip 0.1in
Consider accompanying the $SO(17,1)$ transformation described above
with a lattice boost that decreases the size of the interval
\cite{ginine}:
\begin{equation}
 e^{-\alpha_{00}} = {{1}\over{1+|{\bf A}|^2/4}} \quad .
\label{eq:scale}
\end{equation}
This recovers the Spin(32)/Z$_2$ theory compactified on an interval
of size $2/x$, but with Wilson line ${\bf A}$$=$$x{\rm diag}(1^8,0^8)$
\cite{ginine}. Thus, taking the large radius limit in the {\em dual}
variable, and with dual Wilson line background, yields instead the
spacetime supersymmetric 10D $Spin(32)/Z_2$
heterotic string.
It follows that the $E_8$$\times$$E_8$
and $SO(32)$ heterotic strings share the same
tachyon-free finite temperature ground state with gauge symmetry
$SO(16)$$\times$$SO(16)$. The Kosterlitz-Thouless
transformation at $T_C$$=$$1/\pi\alpha^{\prime 1/2}$ is a self-dual
continuous phase transition in this theory.

\vskip 0.1in
The thermal self-duality transition at the Kosterlitz-Thouless point
in this theory is in the same universality class as that in
the closed bosonic string: namely, the partial derivatives of
the free energy to arbitrary order are all continuous functions
of temperature. An important
distinction is that the string vacuum functional, the
Helmholtz and Gibbs free energies, the internal energy,
and all subsequent thermodynamic potentials, are both finite and
tachyon-free. Finally, the expression for the entropy of the
heterotic string gas has no analog of the unphysical
fixed point entropy present in the bosonic string gas. This
is a consequence of the action of the superorbifold group.

\subsection{Holography and the Self-Dual Phase Transition}

\vskip 0.1in
The generating functional for connected one-loop vacuum string
graphs is invariant under the thermal duality transformation:
$W(T)$ $=$ $W(T^2_c/T)$,
with self-dual temperature, $T_c$ $=$ $1/\pi \alpha^{\prime 1/2}$.
As pointed out by Polchinski in the case of the bosonic
string gas \cite{polchinskibook}, we can infer
the following thermal duality relation which holds for both the
Helmholtz free energy,
$F(T)$ $=$ $-T \cdot W(T)$, and the effective potential,
$\rho(T)$ $=$ $-T \cdot W(T)/V$:
\begin{equation}
F(T)  = {{T^2}\over{T_C^2}} F({{T_C^2}\over{T}}) ,
\quad \quad \rho(T) = {{T^2}\over{T^2_C }} \rho({{T_C^2}\over{T}})
\quad .
\label{eq:thermi}
\end{equation}
Consider the high temperature limit of this expression:
\begin{equation}
\lim_{T \to \infty } \rho(T) =
\lim_{T \to \infty}
{{T^2}\over{T^2_C }} \rho({{T_C^2}\over{T}})
 =  \lim_{(T_C^2/T^2) \to 0} {{T^2}\over{T^2_C }}
\rho({{T_C^2}\over{T}})
 =  {{T^2}\over{T^2_C }} \rho_0
\quad ,
\label{eq:thermasy}
\end{equation}
where $\rho_0$ is the cosmological constant, or vacuum energy
density, at zero temperature. Likewise, at high temperatures,
the free energy grows as the square of the temperature. Thus,
the growth in the number of degrees of freedom at high temperature
in the gas of free heterotic strings is only as fast as in
a {\em two-dimensional} field theory. This is significantly
slower than the $T^{10}$ growth of the high temperature degrees
of freedom in the finite temperature field theory limit.
We comment that the coincidence of the self-dual point of the
orbifold fixed line with the Hagedorn point of the heterotic string
implies that holographic duality holds at the Hagedorn point---
as originally conjectured in \cite{aw}.

\vskip 0.1in
Notice that the prefactor in the high temperature relation is
unambiguous, a consequence of the normalizability of the generating
functional of one-loop vacuum graphs in string theory
\cite{poltorus}. The pre-factor, $\rho_0/T_C^2$, is also background
dependent: it can be computed as a continuously varying function of
the background fields upon compactification to lower spacetime
dimension \cite{gv}. The relation in Eq.\ (\ref{eq:thermasy}) is
unambiguous evidence of the holographic nature of perturbative
{\em closed} string
theory: a reduction in the degrees of freedom in free string theory
at high temperatures, or at short distances \cite{aw}.

\vskip 0.1in
Starting with the duality invariant expression for the string
effective action functional, we can derive the thermodynamic
potentials of the free heterotic string gas.
The Helmholtz free energy of the tachyon-free gas of free heterotic
strings follows from the definition below Eq.\ (\ref{eq:het}), and
is clearly finite at $T_H$, with no evidence for either divergence
or discontinuity.
The internal energy of the free
heterotic string gas takes the form:
\begin{equation}
 U = - \left ( {{\partial W}\over{\partial \beta }} \right )_V =
 \half \int_{\cal F}
{{|d\tau|^2}\over{4\tau_2^2}} (2\pi\tau_2)^{-9/2}
   |\eta(\tau)|^{-16}
{{4\pi \tau_2}\over{\beta}}
\sum_{n,w \in {\rm Z} }
  \left ( {{w^2 x^2}\over{4}} - {{n^2}\over{x^2}}  \right )
\cdot q^{ {{1}\over{2}}{\bf l}_L^2 }
     {\bar q}^{{{1}\over{2}}{\bf l}_R^2 }
\cdot Z_{\rm [SO(16)]^2} \quad ,
\label{eq:term}
\end{equation}
where $Z_{\rm [SO(16)]^2}$ denotes the sums over spin structures
appearing in Eq.\ (\ref{eq:iden}).
$U(\beta)$ vanishes precisely at the self-dual temperature,
$T_c$$=$$1/\pi\alpha^{\prime 1/2}$, $x_c$$=$${\sqrt{2}}$, where the
internal energy contributed by winding sectors cancels that
contributed by momentum sectors. The internal energy changes sign at
$T$ $=$ $T_C$, transitioning from negative values at low temperature
to {\em positive} values at high temperature.

\vskip 0.1in
Referring to the definitions in Eq.\ (\ref{eq:freene}),
the pressure of the free heterotic string gas is clearly zero.
It is easy to demonstrate the analyticity of infinitely many
thermodynamic potentials in the vicinity of the critical point.
It is convenient to define:
\begin{equation}
[d \tau ] \equiv \half
\left [ {{|d\tau|^2}\over{4\pi\tau_2^2}} (2\pi\tau_2)^{-9/2}
   |\eta(\tau)|^{-16}
\cdot Z_{\rm [SO(16)]^2}
e^{2\pi i nw \tau_1} \right ] ,
\quad y(\tau_2;x)
   \equiv 2\pi \tau_2 \left ( {{n^2}\over{x^2}} + {{w^2 x^2}\over{4}}
  \right ) \quad .
\label{eq:vars}
\end{equation}
Denoting the $m$th partial derivative with
respect to $\beta$ at fixed volume by
$W_{(m)}$, $y_{(m)}$,
we note that the higher derivatives of
the generating functional
take the simple form:
\begin{eqnarray}
W_{(1)} =&& \sum_{n,w} \int_{\cal F} [d\tau] e^{-y} (-y_{(1)})
\cr
W_{(2)} =&&
 \sum_{n,w} \int_{\cal F} [d\tau] e^{-y}
(-y_{(2)} + (-y_{(1)})^2 )
\cr
W_{(3)} =&& \sum_{n,w} \int_{\cal F} [d\tau] e^{-y}
(-y_{(3)} - y_{(1)} y_{(2)} +  (-y_{(1)})^3 )
\cr
 \cdots =&& \cdots
\cr
W_{(m)} =&& \sum_{n,w} \int_{\cal F} [d\tau] e^{-y}
(-y_{(m)} - \cdots +  (-y_{(1)})^m )
\quad .
\label{eq:effm}
\end{eqnarray}
Referring back to the definition of $y$, it is easy to see that the
generating functional and, consequently, the full set of thermodynamic
potentials is analytic in $x$. Notice that third and higher derivatives
of $y$ are determined by the momentum modes alone:
\begin{equation}
y_{(m)} = (-1)^m n^2 {{(m+1)! }\over{x^{m+2}}} , \quad m \ge 3 \quad .
\label{eq:ders}
\end{equation}
For completeness, we give explicit results for the first few
thermodynamic potentials:
\begin{equation}
F = - {{1}\over{\beta}} W_{(0)} , \quad
U = - W_{(1)} , \quad
S = W_{(0)} - \beta W_{(1)} , \quad
C_V = \beta^2 W_{(2)} , \cdots \quad .
\label{eq:thermodl}
\end{equation}
Let $2\pi \tau_2$ $\equiv$ $t$.
The entropy is given by the expression:
\begin{equation}
S = \sum_{n,w} \int_{\cal F} [d\tau] e^{-y}
  \left [ 1 + 2 t ( - {{n^2}\over{x^2}} + {{w^2 x^2}\over{4}} )
\right ] \quad ,
\label{eq:entropy1}
\end{equation}
For the specific heat at constant volume, we have:
\begin{equation}
C_V =  \sum_{n,w} \int_{\cal F} [d\tau] e^{-y}
  \left [   4 t^2 ( - {{n^2}\over{x^2}} + {{w^2 x^2}\over{4}} )^2
   - 2t ( 3 {{n^2}\over{x^2}} + {{ w^2 x^2}\over{4}} )
\right ] .
\label{eq:spc}
\end{equation}
It is easy to check the sign of the specific heat at criticality
for an infrared finite self-dual ensemble of closed strings. Since
the first term in square brackets will not contribute at $T_C$,
a stable thermodynamic ensemble with positive specific heat
requires {\em positive} Helmholtz free energy.
This is due to the excess of spacetime fermions over spacetime bosons in the
physical state spectrum.

\vskip 0.1in
Explicitly, the Helmholtz
free energy takes the form:
\begin{equation}
F (\beta) = - \half {{1}\over{\beta}} \int_{\cal F}
{{|d\tau|^2}\over{4\pi\tau_2^2}} (2\pi\tau_2)^{-9/2}
   |\eta(\tau)|^{-16}
\left [ \sum_{n,w}
Z_{\rm [SO(16)]^2}
q^{ {{1}\over{2}}{{\alpha' p_L^2}\over{2}} }
    {\bar q}^{{{1}\over{2}} {{\alpha' p_R^2}\over{2}}}
\right ]
\quad ,
\label{eq:freeee}
\end{equation}
while for the entropy of the free heterotic string gas,
we have the result:
\begin{equation}
S (\beta) =
\half \int_{\cal F}
{{|d\tau|^2}\over{4\pi\tau_2^2}} (2\pi\tau_2)^{-9/2}
   |\eta(\tau)|^{-16}
\sum_{n,w} \left [
   1 + 4 \pi \tau_2 ( - {{n^2}\over{x^2}} + {{w^2 x^2}\over{4}} )
\right ]
Z_{\rm [SO(16)]^2}
q^{ {{1}\over{2}}{{\alpha' p_L^2}\over{2}} }
    {\bar q}^{{{1}\over{2}} {{\alpha' p_R^2}\over{2}}} \quad .
\label{eq:entropy}
\end{equation}
Notice that, unlike the pedagogical, but also unphysical, case of
the bosonic string, there is {\em no analog of fixed point entropy} in
the expression for the entropy of the free heterotic string gas.
Most importantly,
the thermodynamic potentials of the
heterotic ensemble are finite normalizable functions
at all temperatures starting from zero.

\vskip 0.1in
In summary,
the free heterotic string gas displays a continuous
phase transition at the self-dual temperature, unambiguously
identifying a phase transition of the Kosterlitz-Thouless
type \cite{kosterlitz,bosonic}.
We emphasize that
the expressions for the Helmholtz free energy
and generic thermodynamic potentials are {\em not} invariant
under thermal duality transformations. The only thermal
duality invariant quantity is the generating functional for
vacuum string graphs.

\vskip 0.1in
As mentioned above, the sum over spin structures,
$Z_{\rm [SO(16)]^2}$, of the
$SO(16)$$\times$$SO(16)$ string
is known to be {\em negative} \cite{dh,agmv,itoyama,gv}. This
is due to the preponderance of spacetime fermionic modes over
spacetime bosonic modes. The internal energy of free
heterotic strings is therefore negative
at low temperatures, vanishing precisely at the self-dual
temperature $x_c$$=$$2^{1/2}$,
$T_c$$=$$1/\pi \alpha^{\prime 1/2}$.
The Helmholtz and Gibbs free energies are minimized at
the self-dual temperature, and the specific heat at
constant volume is positive.

\section{The Free Type I Open and Closed String Gas}

\subsection{Cancellation of the Dilaton Tadpole}

\vskip 0.1in
In the previous section, we have seen that the ten-dimensional supersymmetric
$E_8$$\times$$E_8$ and $SO(32)$ heterotic string theories have a unique tachyon-free
ground state at finite temperature in the presence of a temperature dependent timelike
Wilson line. The gauge group is $SO(16)$$\times$$SO(16)$. In nine dimensions
and below, the two different heterotic string theories are equivalent upto a target
space duality. The strong coupling dual of this theory is the type IB unoriented string
theory with identical gauge group. We can obtain limited insight
into the strong coupling behaviour of the heterotic string gas by
analyzing a gas of free type I open and closed strings.

\vskip 0.1in
It will be convenient to work in the timelik
T$_0$-dual type I$^{\prime}$ picture with 16
D8branes on an orientifold plane, each identified with its own image-brane.
We turn on the temperature-dependent timelike Wilson line,
$A_0$$=$${{1}\over{\beta}}((1)^8,0^8)$,
thereby partitioning the D8branes into two sets. We will label
open strings connecting two D8branes in the same partition by
the index $w$$=$$0$. At finite temperature, this gives a total of
($2$$\cdot$$8$$\cdot$$7$)$\cdot$$2$ massless states from zero length
open strings connecting a pair of D8branes within each of two partitions.
In addition, there are $8$$\cdot$$2$
photons from zero length open strings connecting each D8brane to itself. Open strings
which connect branes in distinct partitions are labelled $w$$=$$1$. They are
ordinarily massive, resulting in a breaking of the gauge group from
$O(32)$ to $O(16)$$\times$$O(16)$ at finite temperature.

\vskip 0.1in
Consider the free energy, $F(\beta)$, of a gas of free type I$^{\prime}$ strings
in this ground state. $F$ is obtained from the generating
functional for connected vacuum string graphs, $\beta F(\beta)$$=$$- W(\beta)$. The
one-loop contribution to $W$
is free of dependence on the string coupling and is given by the Polyakov
path integral summing surfaces with two boundaries, with
a boundary and a crosscap, or with two
crosscaps \cite{polchinskibook}. The boundaries and crosscaps are localized on the orientifold
plane but the worldsheet itself extends into the transverse Dirichlet directions. We will
use the worldsheet formalism described in \cite{polchinskibook}.
$W(\beta)|_{\beta =\infty}$ is required to
agree with the vacuum functional of the supersymmetric $O(32)$ type I$^{\prime}$ string
at zero temperature. We also require a tachyon-free thermal spectrum
which retains the good ultraviolet and infrared behavior of the zero temperature
spectrum. Thus, the {\em independent} cancellations of dilaton and Ramond-Ramond ten-form
tadpoles must continue to hold at finite temperature precisely as at zero temperature
\cite{polchinskibook}. The ten-form potential has vanishing eleven-form field strength and
there is no propagator for this field; the tadpole must therefore be set to zero both at
zero and at finite temperature. As a consequence, we have no choice but to set the
dilaton tadpole to zero--- even in the absence of spacetime supersymmetry, since the
remaining conditions on $F(\beta)$ are too restrictive.

\vskip 0.1in
On the other hand, the usual cancellation between closed string NS-NS and R-R exchanges
resulting from spacetime supersymmetry must not hold except in the zero temperature limit.
This is in accordance with the finite temperature analysis for closed string theories
and is achieved by introducing temperature dependent phases
\cite{aw}. We will insert an identical temperature-dependent phase for the
contributions to $F$ from each of the worldsheet topologies in order to preserve the
zero temperature tadpole cancellations. The result for the free energy with $N$
D8branes on the orientifold plane takes the {\em unique} form:
\begin{eqnarray}
F = && \beta^{-1} \int_0^{\infty} {{dt}\over{8t}}
  {{(2\pi t )^{-9/2}}\over{
  \eta(it)^{8}}}
[ ~  2^{-8} N^2 ( Z_{\rm [0]} - e^{\i \pi n } Z_{\rm [1]})
 \cr
\quad && \quad\quad
+   Z_{\rm [0]} - e^{\i \pi n } Z_{\rm [1]}
 ~ - ~ 2^{-3} N  Z_{\rm  [0]} \cr
\quad&& \quad\quad\quad + ~ 2^{-3} N e^{\i \pi n } Z_{\rm [1]} ~ ]
\times \sum_{n \in {\rm Z}; w=0,1 } q^{\alpha^{\prime}\pi^2 n^2
/\beta^2 +
   w^2 \beta^2/ 4 \pi^2 \alpha^{\prime} }
\label{eq:freeIp}
\end{eqnarray}
The Matsubara frequency spectrum is given by the eigenvalues for timelike
momentum: $p_0$$=$$2n\pi/\beta$, with $n$$\in$${\rm Z}$. Notice that $w$
only takes the values $0$, $1$. Unlike the closed string gases described
in previous sections,
the spectrum breaks both modular invariance and also, as a
consequence, thermal duality invariance. The $n$$>$$1$ modes have
no $w$$>$$1$ mode counterparts in the open string thermal spectrum.

\vskip 0.1in
Are there any new tadpoles
in the expression for $W$ due to finite temperature excitations? We find none. The variable
$q$$=$$e^{-2\pi t}$ is the modular parameter for the cylinder, and we have used the
identifications: $2t_K$$=$$t_C$, and $2t_M$$=$$t_C$, to express the Mobius strip and
Klein bottle amplitudes in terms of the cylinder's modular parameter. As in
\cite{polchinskibook}, the subscripts $[0]$, $[1]$, denote, respectively, NS-NS and R-R closed
string exchanges. Thus,
\begin{eqnarray}
Z_{\rm [0]} =&& ({{\Theta_{00}(it;0)}\over{\eta(it)}})^4 -
({{\Theta_{10}(it;0)}\over{\eta(it)}})^4
\cr
Z_{\rm [1]} =&& ({{\Theta_{01}(it;0)}\over{\eta(it)}})^4 -
({{\Theta_{11}(it;0)}\over{\eta(it)}})^4
\label{eq:relats}
\end{eqnarray}
where the (00), (10), (01), and (11), denote, respectively, (NS-NS), (R-NS),
(NS-R), and (R-R), boundary conditions on worldsheet fermions in the closed string sector.
The boundary conditions in the open string sectors can be read off using worldsheet
duality \cite{polchinskibook}. As in the supersymmetric zero temperature limit, the
independent cancellation of both the dilaton and the Ramond-Ramond ten-form tadpole
requires that $N$$=$$16$.

\vskip 0.1in
Notice that, unlike the
case of the closed string gases, despite the fact that spacetime
supersymmetry has been broken the free energy nevertheless vanishes at this order
in string perturbation theory! By factorization of the one-loop amplitude, the absence
of all tadpole graphs will imply the vanishing of the one-loop free energy \cite{polchinskibook}.
It is tempting to interpret our result as indication that the weakly coupled heterotic
$SO(16)$$\times$$SO(16)$ ground state with nonvanishing dilaton tadpole--- as inferred from
the nonvanishing oneloop cosmological constant obtained in the previous section, \lq\lq flows"
to strong coupling. Fortunately, this strongly coupled heterotic ground state with
vanishing dilaton one-point function also has a dual weakly coupled type I$^{\prime}$
string description given here. We comment that while such an inference must remain
conjectural the physical conclusions it leads to are satisfactory. We should note that
the weak-strong duality between heterotic and type I string theories in the absence
of supersymmetry has been studied in \cite{dienes}.

\vskip 0.1in
We close with an important observation. Recall that the temperature
dependent timelike Wilson line contributes a shift in the mass formula,
$\Delta p_0$$=$$x$, thus removing all of the potential low temperature
tachyons arising from $n$$\neq$$0$ momentum modes. The $n$$=$$0$,
$w$$=$$1$ mode is ordinarily massive, turning massless only in the
zero temperature limit of coincident branes and enhanced
gauge symmetry. Given the special role played by the $(1,0)$ and $(0,1)$
thermal modes in closed string gases, one might wonder at the role of the
$n$$=$$1$, $w$$=$$0$ mode in the type I$^{\prime}$ gas. We will find that
the $(1,0)$ mode signals a phase transition to a high temperature long
string phase. The phase transition occurs at a
critical temperature at, or below, $T_C$.

\subsection{Transition to the High Temperature Long String Phase}

\vskip 0.1in
As mentioned earlier, the expression for the generating functional of one-loop
vacuum string graphs given in Eq.\ (\ref{eq:freeIp}) explicitly violates thermal
duality and, unlike the case of closed string gases, there is no thermal
self-duality transition at $T_C$. Instead, at a critical temperature at, or below,
$T_C$, we will find concrete evidence of a phase transition between a low temperature
phase of short open strings and a high temperature long string phase.
As in the case of closed strings, the phase transition belongs
in the Kosterlitz-Thouless universality class. The thermodynamic potentials of
the free open and closed string gas are analytic functions of temperature at
criticality.

\vskip 0.1in
The calculation that follows relies on the well-known fact that the
sub-string-scale dynamics of string theory can be probed by D0branes: pointlike
topological string solitons whose mass scales as $1/g$, in the presence of
external background fields \cite{dbrane,dkps,flux}. In the free string limit the
pointlike D0branes behave like analogs of infinitely massive heavy quarks:
semiclassical, theoretical probes of the confinement regime. It is well known
that an order parameter signalling the thermal deconfinement phase transition in a
nonabelian gauge theory is the expectation value of a closed timelike Wilson
loop. We therefore look for a signal of a thermal phase transition
in the high temperature asymptotics of the pair correlator of
timelike Wilson loops in finite temperature string theory. We will
use the fact that
the pair correlator has a simple worldsheet representation \cite{cmnp,pair,flux}.

\vskip 0.1in
The pair correlator of closed timelike Wilson loops computes the Minkowskian
time propagator of a D0brane pair with fixed spatial separation $r$. The loops
lie within the worldvolume of the D8brane stack on a single orientifold plane.
In \cite{pair,flux} we found that in the presence of an external electric field,
${\cal F}_{09}$, and quantized background ${\cal B}_{[2]}$ and ${\cal C}_{[0]}$
fields, the D0branes experience an attractive binding potential of the form
$u^4\alpha^{\prime 4}/r^9$ at short distances and at zero temperature.
The dimensionless parameter $u$ is defined as $u$$=$${\rm tanh}^{-1} {\cal F}^{09}$.
In the $T_9$-dual type I$^{\prime}$ picture, $u$ has a dual interpretation as the
relative velocity of the D0branes in a direction transverse to their separation.
A systematic expansion in field strength to all orders gives the small velocity,
short distance, corrections to the leading field-dependent binding interation
between D0branes \cite{pair,flux}.

\vskip 0.1in
We consider parallel and oriented timelike Wilson loops in the D8brane stack wrapped
around the Euclidean coordinate $X^0$, and with fixed spatial separation $r$. They
represent the Euclidean worldlines of a pair of static and closely separated semiclassical
D0brane sources. The path integral expression for the pair correlation function,
${\cal W}_2$, at finite temperature takes the form \cite{cmnp,pair,flux,decon}:
\begin{eqnarray}
{\cal W}_2(r,\beta)  =&& \lim_{r \to 0}
\int_0^{\infty} dt {{e^{- r^2 t/2\pi \alpha^{\prime} }}\over{\eta(it)^{8}}}
       \sum_{n\in {\rm Z}; w=0,1} q^{\pi^2 n^2 \alpha^{\prime}/\beta^2
        + w^2 \beta^2 / 4 \pi^2 \alpha^{\prime} }
\cr
\quad && \quad \times [~ ({{\Theta_{00}(it;0)}\over{\eta(it)}})^4
 -  ({{\Theta_{10}(it;0)}\over{\eta(it)}})^4  \cr
\quad &&\quad - e^{\i \pi n}
\{ ({{\Theta_{01}(it;0)}\over{\eta(it)}})^4 -
    ({{\Theta_{11}(it;0)}\over{\eta(it)}})^4 \} ~] .
\label{eq:pairc}
\end{eqnarray}
The static pair potential at short distances is extracted from the dimensionless
amplitude ${\cal W}_2$ as follows. We set
${\cal W}_2$$=$$\lim_{\tau\to\infty} \int_{-\tau}^{+\tau} d\tau V[r(\tau),\beta]$,
inverting this relation to express $V[r,\beta]$ as an integral over the modular
parameter $t$. Consider a $q$ expansion of the integrand, valid in the limit
$r$$\le$$2\pi \alpha^{\prime}$, $t$$\to$$\infty$, where the shortest open strings
dominate the modular integral.
Retaining the leading terms in the $q$ expansion and performing explicit term-by-term
integration over the worldsheet modulus, $t$, \cite{pair,flux}, isolates the following
short distance interaction \cite{decon}:
\begin{eqnarray}
V(r,\beta) =&&  (8\pi^2 \alpha^{\prime})^{-1/2}
\int_0^{\infty} dt e^{- r^2 t/2\pi \alpha^{\prime} } t^{1/2}
\cr
\quad &&\quad \times \sum_{n\in {\rm Z}}
( 16 - 16 e^{i \pi n } ) q^{\alpha^{\prime} \pi^2 n^2 /\beta^2} + \cdots
\cr
\quad =&&2^4 {{1}\over{r(1+ {{
  r_{\rm min.}^2 }\over{r^2}} {{\beta_C^2 }\over{\beta^2 }} )^{1/2} }} + \cdots
\label{eq:static}
\end{eqnarray}
where we have dropped all but the contribution from the $n$$=$$1$ thermal mode in
the last step. We have expressed the result in terms of the
characteristic minimum distance scale probed in the absence of
external fields, $r_{\rm min.}$$=$$2\pi \alpha^{\prime 1/2}$, and the
closed string's self-dual temperature $T_C$.
Consider the crossover in the behavior of the interaction as a function of
temperature:
at low temperatures, with $(r/r_{\rm min.})\beta$$>>$$\beta_C$,
we can expand in a power series. The leading correction to the inverse
power law is $O(1/\beta^2 r^3)$. At high temperatures with
$(r/r_{\rm min.}) \beta$$<<$$\beta_C$, the potential instead approaches a
constant independent of $r$. We note the characteristic signal of the
onset of the (long string) confining phase where the order parameter approaches
a constant independent of $r$. Notice that at the
crossover between the two phases the potential is a precise
inverse power law. Note also that the overall coefficient
of the potential is independent of the spacetime dimensionality of the gauge
fields: the factor of $2^4$ is related to the critical dimension of the
type I$^{\prime}$ string--- not the dimensionality of the Dpbrane in
question \cite{polchinskibook}.

\vskip 0.1in In the presence of an external electromagnetic field,
the result given above will be modified by the simple replacement,
$r_{\rm min.}$$\to$$2\pi\alpha^{\prime 1/2}u$, where ${\cal
F}^{09}$$=$${\rm tanh}^{-1}u$ is the electric field strength. The
transition temperature, $T_d$$=$$u T_C$, is consequently {\em
lower} than the closed string's self-dual temperature. Thus, a
background with a flux tube of size roughly a hundredth of
$\alpha^{\prime 1/2}$ can give a suppression of $10^{-3}$ in the
transition temperature. It is important to note that $r$ cannot be
taken {\em arbitrarily} small since the path integral
representation for $V(r)$ itself breaks down--- the integral
becomes formally divergent.

\vskip 0.1in
It is interesting to compare these results with an old conjecture in the gauge theory
literature. Peskin has argued \cite{peskin} that when the deconfining phase transition
in a renormalizable gauge theory is second order or is, more generally, a continuous
phase transition without discontinuity in the free energy, the heavy quark potential
must necessarily take the scale-invariant form, $C/r$, with $C$ a constant independent
of spacetime dimension. Our string theory calculation is a beautiful illustration of
this phenomenon--- although in an entirely different context.
Perturbative string theory is fully renormalizable, and its low energy limit is a
renormalizable gauge theory. The phase transition we have
observed is a continuous phase transition. The D0brane pair potential takes the
scale-invariant $-1/r$ form at criticality. And the coefficient of the potential is
independent of the spacetime dimension of the gauge theory.

\subsection{Scaling Relations and the Thermodynamic Potentials}

\vskip 0.1in
We will now examine the behavior of the thermodynamics potentials of the free
open and closed string gas. Our first goal is to point out that the scaling
relations for the free energy of the open and closed string gas differ in the
low and high temperature phases. The low temperature behavior of $F$ is determined
by the $n$$=$$1$, $w$$=$$0$ mode. The integral expression for the free energy
is dominated by the $t$$\to$$\infty$ limit with short open strings. Isolating
this limit gives:
\begin{eqnarray}
\lim_{\beta \to \infty} F = && \lim_{\beta \to \infty} \beta^{-1} \int_0^{\infty}
{{dt}\over{8t}}
  {{(2\pi t )^{-9/2}}\over{
  \eta(it)^{8}}}
[ ~  2^{-8} N^2 ( Z_{\rm [0]} - e^{\i \pi n } Z_{\rm [1]})
 \cr
\quad && \quad\quad
+   Z_{\rm [0]} - e^{\i \pi n } Z_{\rm [1]}
 ~ - ~ 2^{-3} N  Z_{\rm  [0]} \cr
\quad&& \quad\quad\quad + ~ 2^{-3} N e^{\i \pi n } Z_{\rm [1]} ~ ]
\times \sum_{n \in {\rm Z}, w=0,1 } q^{\alpha^{\prime}\pi^2 n^2
/\beta^2 +
    w^2 \beta^2/4\pi^2 \alpha^{\prime} }
\cr
 =&& \beta^{-1} \left ( \beta^{-9} \cdot \rho_{\rm low} \right ) \quad ,
\label{eq:freeIpp}
\end{eqnarray}
where $\rho_{\rm low}$ is a constant independent of temperature. Of course
$\rho_{\rm low}$ vanishes since the free energy itself vanishes owing to the
cancellation of dilaton tadpoles. But our interest here is in the scaling
behavior of the free energy. Not surprisingly, we find the expected $T^{10}$
growth of the degrees of freedom in the ten-dimensional finite temperature
gauge theory. This is the low energy limit of the type I string picked out
by the $t$$\to$$\infty$ asymptotics.

\vskip 0.1in
The high temperature behavior of $F(\beta)$ is determined instead by the $n$$=0$,
$w$$=$$1$ thermal mode, and the dominant contribution to the integral is
from the $t$$\to$$0$ limit. A modular transformation, $t$$\to$$1/t$,
puts the integrand in a suitable form for a term-by-term expansion. Isolating the
$t$$\to$$0$ asymptotics as usual \cite{polchinskibook},
we can perform the $t$ integral to obtain a most unexpected result:
\begin{eqnarray}
\lim_{\beta \to 0} F = && \lim_{\beta \to 0} \beta^{-1} \int_0^{\infty}
{{dt}\over{8t}}
  {{(2\pi t )^{-3/2}}\over{
  \eta(i/t)^{8}}}
[ ~  2^{-8} N^2 ( Z_{\rm [0]} - e^{\i \pi n } Z_{\rm [1]})
 \cr
\quad && \quad\quad
+   Z_{\rm [0]} - e^{\i \pi n } Z_{\rm [1]}
 ~ - ~ 2^{-3} N  Z_{\rm  [0]} \cr
\quad&& \quad\quad\quad + ~ 2^{-3} N e^{\i \pi n } Z_{\rm [1]} ~ ]
\times \sum_{n \in {\rm Z}, w=0,1 } q^{\alpha^{\prime}\pi^2 n^2
/\beta^2 +
    w^2 \beta^2/4\pi^2 \alpha^{\prime} }
\cr
 =&& \beta^{-1} \left ( \beta^3 \cdot \rho_{\rm high}  \right ) \quad ,
\label{eq:freeIpph}
\end{eqnarray}
where $\rho_{\rm high}$ is a constant independent of temperature.
The {\em lowering} of the free energy with an increase in temperature
at very high temperatures can be understood as the shift in balance
between internal energy and entropy in the production of the long
string. Recall that $F$$=$$U$$-$$TS$, so that although the long string
grows at a cost in the internal energy--- which also scales as $\beta^2$
at high temperatures (see below), this growth is entropically favoured,
showing up as a deficit in the free energy.
We remind the reader that this analysis
has neglected gravitational self-interactions which will eventually be
the determining factor in the fate of the long string phase
\cite{sussk,garyp}.

\vskip 0.1in
It is helpful to verify the corresponding scaling relations for the first
few thermodynamic potentials. The internal energy of the free type I
string gas takes the form:
\begin{equation}
 U = - \left ( {{\partial W}\over{\partial \beta }} \right )_V =
 \half \int_{0}^{\infty}
{{dt}\over{8t}} {{(2\pi t )^{-9/2}}\over{
  \eta(it)^{8}}} \cdot Z_{\rm open}
\sum_{n \in {\rm Z}, w=0,1 }
{{4 \pi t}\over{\beta}}  \left (
{{w^2 \beta^2}\over{4\pi^2 \alpha^{\prime}}} -
   {{\alpha^{\prime} \pi^2 n^2}\over{\beta^2}}  \right )
q^{\alpha^{\prime}\pi^2 n^2 /\beta^2 +
    w^2 \beta^2/4\pi^2 \alpha^{\prime} }
 \quad ,
\label{eq:entermn}
\end{equation}
where ${\rm Z}_{\rm open}$ is the factor in square brackets in
the expression in Eq.\ (\ref{eq:freeIp}). Unlike the heterotic string gas,
$U(\beta)$ no longer vanishes at the self-dual temperature,
$T_c$$=$$1/\pi\alpha^{\prime 1/2}$, since there are no $w$$>$$1$
modes corresponding to the $n$$>$$1$ modes.
The internal energy is ordinarily negative except at very
high temperatures where the $w$$=$$1$ mode dominates.

\vskip 0.1in
Referring to the definitions in Eq.\ (\ref{eq:freene}),
the pressure of the free type I string gas is zero.
The analyticity of infinitely many
thermodynamic potentials in the vicinity of the critical point
can be demonstrated as for the heterotic string gas.
We define:
\begin{equation}
[d t ] \equiv \half
\left [
{{dt}\over{8t}} {{(2\pi t )^{-9/2}}\over{
  \eta(it)^{8}}} \cdot Z_{\rm open}
\right ] ,
\quad y(t;\beta)
   \equiv 2\pi t \left (
{{w^2 \beta^2}\over{4\pi^2 \alpha^{\prime}}} -
   {{\alpha^{\prime} \pi^2 n^2}\over{\beta^2}}
  \right ) \quad ,
\label{eq:varsn}
\end{equation}
and denote the $m$th partial derivative with
respect to $\beta$ at fixed volume by
$W_{(m)}$, $y_{(m)}$.
The higher derivatives of
the generating functional
now take the simple form:
\begin{eqnarray}
W_{(1)} =&& \sum_{n,w} \int_{0}^{\infty} [dt] e^{-y} (-y_{(1)})
\cr
W_{(2)} =&&
 \sum_{n,w} \int_{0}^{\infty} [dt] e^{-y}
(-y_{(2)} + (-y_{(1)})^2 )
\cr
W_{(3)} =&& \sum_{n,w} \int_{0}^{\infty} [dt] e^{-y}
(-y_{(3)} - y_{(1)} y_{(2)} +  (-y_{(1)})^3 )
\cr
 \cdots =&& \cdots
\cr
W_{(m)} =&& \sum_{n,w} \int_{0}^{\infty} [dt] e^{-y}
(-y_{(m)} - \cdots +  (-y_{(1)})^m )
\quad .
\label{eq:effmn}
\end{eqnarray}
Referring back to the definition of $y$, it is easy to see that the
generating functional and, consequently, the full set of thermodynamic
potentials is analytic in $\beta$. Notice that third and higher derivatives
of $y$ are determined by the momentum modes alone:
\begin{equation}
y_{(m)} = (-1)^m n^2 {{(m+1)! }\over{\beta^{m+2}}} , \quad m \ge 3 \quad .
\label{eq:dersn}
\end{equation}
Explicit expressions for the first few
thermodynamic potentials are as follows:
\begin{equation}
F = - {{1}\over{\beta}} W_{(0)} , \quad
U = - W_{(1)} , \quad
S = W_{(0)} - \beta W_{(1)} , \quad
C_V = \beta^2 W_{(2)} , \cdots \quad .
\label{eq:thermodln}
\end{equation}
The entropy is given by the expression:
\begin{equation}
S = \sum_{n,w} \int_{0}^{\infty} [dt] e^{-y}
  \left [ 1 + 2 t (
{{w^2 \beta^2}\over{4\pi^2 \alpha^{\prime}}} -
   {{\alpha^{\prime} \pi^2 n^2}\over{\beta^2}} )
\right ] \quad ,
\label{eq:entropy1n}
\end{equation}
For the specific heat at constant volume, we have:
\begin{equation}
C_V =  \sum_{n,w} \int_{0}^{\infty} [dt] e^{-y}
  \left [   4 t^2 (
{{w^2 \beta^2}\over{4\pi^2 \alpha^{\prime}}} -
   {{\alpha^{\prime} \pi^2 n^2}\over{\beta^2}} )
)^2
   - 2t (
{{w^2 \beta^2}\over{4\pi^2 \alpha^{\prime}}} +
   3 {{\alpha^{\prime} \pi^2 n^2}\over{\beta^2}} )
\right ] .
\label{eq:spcn}
\end{equation}
We can check the sign of the specific heat at criticality
as for the closed string gas. Since the first term in square
brackets will not contribute at $T_C$,
a stable thermodynamic ensemble with positive specific heat
requires {\em positive} Helmholtz free energy.
As in the case of the heterotic string gas, we have
an excess of spacetime fermions over spacetime bosons in the
physical state spectrum.
The entropy takes the simple form:
\begin{equation}
S (\beta) =
\half \int_{0}^{\infty}
{{dt}\over{8t}} {{(2\pi t )^{-9/2}}\over{
  \eta(it)^{8}}} \cdot Z_{\rm open}
\sum_{n \in {\rm Z}, w=0,1 }
\left [ 1 + 4 \pi t (
{{w^2 \beta^2}\over{4\pi^2 \alpha^{\prime}}} -
   {{\alpha^{\prime} \pi^2 n^2}\over{\beta^2}}  )
 \right ]
q^{\alpha^{\prime}\pi^2 n^2 /\beta^2 +
    w^2 \beta^2/4\pi^2 \alpha^{\prime} }
\quad .
\label{eq:entropynn}
\end{equation}
In summary, the free type I string gas displays a continuous
phase transition at the self-dual temperature
in the universality class of the Kosterlitz-Thouless
transition \cite{kosterlitz,bosonic}. Unlike the case of
the self-dual heterotic string gas where the free energy
satisfies a holographic self-duality relation, here the
free energy scales differently in the low and high
temperature phases. At low temperatures, we recover the
$T^{10}$ growth in the number of degrees of freedom of the
low energy gauge theory limit. The internal energy and free
energy scale as $\beta^{-10}$ at low temperatures, and the
specific heat and entropy scale as $\beta^{-9}$, exactly as
expected in a ten-dimensional quantum field theory at finite
temperature. At high temperatures, we discover an entirely new
long string phase characterized by the free energy and internal
energy scaling as $\beta^{2}$. The entropy and specific heat
scale as $\beta^{3}$.

\section{Conclusions}

\vskip 0.1in
There are many difficulties inherent in developing a satisfactory description
of the statistical mechanics of strings, which can only in part be addressed
within the perturbative worldsheet formalism. We have restricted ourselves to
the free string limit, thus obviating the hurdle of the Jeans instability and
of a nonequilibrium thermodynamics. We emphasize that discussion of these issues
requires a nonperturbative formalism for string theory within which the strong
coupling degrees of freedom are accessible. Proper knowledge of the free string
limit is a prerequisite to such a future analysis. We have attempted to fill
this gap in our work.

\vskip 0.1in
Our results provide concrete evidence of the transition to the long string
phase, long suspected in the string theory literature \cite{polchinskibook,sussk,garyp}.
We identify the phase transition as being in the universality class of the
Kosterlitz-Thouless transition. We clarify that the existence of distinct
low and high temperature
phases is only a feature of the open and closed string gas. The heterotic
string gas undergoes a self-dual phase transition, also in the Kosterlitz-Thouless
universality class, but the low temperature phase is equivalent to the high
temperature phase. Consequently, the heterotic string gas satisfies a
holographic duality relation which
implies that the growth in the number of degrees of freedom at high
temperature equals that in a two-dimensional
quantum field theory. This is not true for the type I string gas. The open and
closed string gas explicitly violates thermal duality, and does not
display holographic scaling relations. The free energy scales as $T^{10}$ at low
temperatures, as expected from a theory with a finite temperature gauge theory
as its low energy limit. The high temperature long string phase has a most
unusual scaling relation with $F(\beta)$$\sim$$T^{-2}$. This behavior is
expected to be modified when the gravitational self-interactions of the long
string are taken into account. That analysis awaits future work.

\vskip 0.1in
It is reassuring that the unphysical features of the free bosonic string gas,
namely, the tachyonic modes and the fixed point entropy, are absent in the
statistical description of the fermionic string gases. In this paper we have
focused on the entropy of a gas of free strings in the trivial flat space
background as well as--- in the case of the open string, in a constant
background field. It is natural to ask whether our calculation can be repeated
for more general string theory backgrounds such as black holes and black strings.
Notice that in these cases there is an additional contribution to the
entropy coming from the classical Polyakov action in the curved background
metric. This term is the intrinsic entropy associated with the background
geometry. Given our understanding of the relation between Dbranes and black hole
entropy, we would expect that repeating the calculations in this paper for
an anomaly-free type I background with a stack of N Dpbrane-anti-Dpbrane pairs
\cite{witsc},
in addition to the 32 D8branes of the O(32) type I string,
should recover the main features of previous results on black hole entropy
\cite{polchinskibook,svm}. It would be most interesting to pursue the details
of such an analysis.

\vspace{0.2in}
\noindent{\bf Acknowledgements:}
I would like to
thank C. Bachas, K. Dienes, A. Dhar, L. Dixon, J. Distler, M. Fukuma,
P. Ginsparg, J. Harvey, G. Horowitz, H. Kawai, I. Kostov, D.
Kutasov, M. Peskin, G. Shiu, E. Silverstein, B. Sundborg,
H. Tye, and E. Witten for comments on an earlier draft.
This research was supported in part by the award of grant
NSF-PHY-9722394 by the National Science Foundation under the
auspices of the Career program.

\vskip 0.5in
\noindent{\bf Note Added (Sep 2005):} Many of the points made in this
paper are either extraneous, or incorrect in the details, although the broad
conclusions summarized in the abstract 
do stand. Namely, that there is no self-consistent type II 
superstring ensemble, in the absence of a Yang-Mills gauge sector.
The fact that both heterotic and type I theory have equilibrium canonical
ensembles free of thermal tachyons; a crucial role is played by the
temperature dependent Wilson line wrapping Euclidean time. An addition from
topics covered in hep-th/0105244 is the discussion of the type IB-I$^{\prime}$
 open and closed string
ensembles, including the
evidence for an order parameter for the unusual duality phase transition in
this theory \cite{decon}. I refer the reader to hep-th/0506143.

\vskip 0.2in
\noindent{\large\bf Appendix}

\vskip 0.1in
For completeness, we gather in this appendix some
relevant facts from the cited literature
on the action of the fermionic orbifold group.
The action of a fermionic R-orbifold group on $\psi^0$, $ X^0$,
consistent with the preservation of an N=1 world-sheet superconformal
invariance, satisfying thermal boundary conditions, and transforming
correctly under a thermal duality transformation, remains to be specified.
To begin with, we recall
the analysis in \cite{dgh} where the distinct possibilities for
fixed lines of N=1 superconformal theories with superconformal central
charge ${\hat c}$ $=$ ${{2}\over{3}} c$ $=$ $1$ were classified. We can
identify points under the reflection $R$,
$ X^0 \simeq - X^0 $, with fundamental region the half-line,
$X^0 \geq 0$, supplementing with the periodic identification,
$t^w : $ $ X^0 \simeq X^0 + 2 \pi w r_{\rm circ.}$, $w \in {\rm Z}$.
The resulting fundamental region is the interval,
$0 \leq X^0 < \pi r_{\rm circ.}$ \cite{polchinskibook}.
An N=1 world-sheet supersymmetry requires
that $\psi^0 \simeq - \psi^0$ under $R$, and the periodic
identification $t^w$ ordinarily preserves the
superconformal generator, $\psi^0 \partial_z X^0$,
leaving the fermions invariant. The partition function of
an Ising fermion is the sum over spin structures (A,A), (A,P),
(P,A), and (P,P), manifestly invariant under $\psi$ $\to$ $-\psi$
\cite{dgh}. Thus, two fixed lines of ${\hat c}$$=$$1$ theories are
obtained by simply
tensoring the Ising partition function with either
${\rm Z}_{\rm circ.}$, or ${\rm Z}_{\rm orb}$, from
the previous section \cite{dgh}:
\begin{equation}
{\rm \hat Z}_{\rm circ.} =
{\rm Z}_{\rm circ.} {\rm Z}_{\rm Ising}, \quad
{\rm \hat Z}_{\rm orb.} =
{\rm Z}_{\rm orb.} {\rm Z}_{\rm Ising} , \quad
{\rm Z}_{\rm Ising} =
\half \left ( |{{\Theta_3}\over{\eta}}| +
 |{{\Theta_4}\over{\eta}}| +
 |{{\Theta_2}\over{\eta}}| \right ) \quad .
\label{eq:is}
\end{equation}
Accompanying $R$, $t^w$ with
the action of the Z$_2$-moded generator,
$(-1)^{N_F}$, where $N_F$ is spacetime fermion number, one
can define additional super-orbifold theories, ${\rm \hat Z}_{\rm sa}$,
${\rm \hat Z}_{\rm so}$, and ${\rm \hat Z}_{\rm orb.'}$, giving a
total of five fixed lines of ${\hat c}$ $=$ $1$ theories \cite{dgh}:
\begin{equation}
{\rm \hat Z}_{\rm orb.'} = R (-1)^{N_F} {\rm \hat Z}_{\rm circ.} , \quad
{\rm \hat Z}_{\rm sa} =  (-1)^{N_F} e^{2\pi i \delta(p)} {\rm \hat Z}_{\rm circ.}, \quad
{\rm \hat Z}_{\rm so} = R {\rm \hat Z}_{\rm sa}
= R (-1)^{N_F} e^{2\pi i \delta(p)} {\rm \hat Z}_{\rm circ.} \quad ,
\label{eq:aff}
\end{equation}
where $\delta(p)$ acts on the momentum lattice as an order two shift in the momentum
eigenvalue, $n$ $\to$ $n+\half$. Note that this shift vector is not thermal
duality invariant. We will adapt the left-right symmetric analysis of \cite{dgh}
for the different ten-dimensional fermionic string theories and taking into
account the thermal boundary conditions. Thus, the action of
the super-orbifold group will be required to preserve either one,
or both, of the holomorphic N=1 superconformal invariances in,
respectively, the heterotic, or type II, string theories, consistent
with modular invariance and the correct thermal duality transformations.

\vskip 0.1in
It is convenient to introduce lattice partition functions that transform
straightforwardly under a thermal duality transformation \cite{dgh}. A
Poisson resummation makes their transformation properties under modular
transformations manifest. As in \cite{bosonic}, we introduce a
dimensionless inverse temperature
(radius) defining $x$ $\equiv$ $r(2/\alpha^{\prime})^{1/2}$,
with $\beta$ $=$ $\pi (\alpha^{\prime}/2)^{1/2} x$. The
dimensionless quantized momenta live in
a $(1,1)$ dimensional Lorentzian self-dual lattice
\cite{nsw,dgh,polchinskibook}:
\begin{equation}
\Lambda^{(1,1)} :  \quad \quad\quad ({{\alpha^{\prime}}\over{2}})^{1/2}
(p_L , p_R) \equiv   (l_L , l_R) = (
{{n}\over{x}} + {{wx}\over{2}}   ,
{{n}\over{x}} - {{wx}\over{2}}  )  \quad ,
\label{eq:dimless}
\end{equation}
with a natural decomposition into even and odd integer momentum
(winding) sums. Thus, we can define:
\begin{eqnarray}
\Gamma^{++} (x) \equiv&& \sum_{w \in 2{\rm Z}, n \in 2{\rm Z}}
 q^{{{1}\over{2}} ({{n}\over{x}} + {{wx}\over{2}} )^2 }
 {\bar q}^{{{1}\over{2}} ({{n}\over{x}} - {{wx}\over{2}} )^2 }
, \quad
\Gamma^{--} (x) \equiv \sum_{w \in 2{\rm Z}+1, n \in 2{\rm Z}+1}
 q^{{{1}\over{2}} ({{n}\over{x}} + {{wx}\over{2}} )^2 }
 {\bar q}^{{{1}\over{2}} ({{n}\over{x}} - {{wx}\over{2}} )^2 }
 \cr
\nonumber
\Gamma^{+-} (x) \equiv&& \sum_{w \in 2{\rm Z}, n \in 2{\rm Z}+1}
 q^{{{1}\over{2}} ({{n}\over{x}} + {{wx}\over{2}} )^2 }
 {\bar q}^{{{1}\over{2}} ({{n}\over{x}} - {{wx}\over{2}} )^2 }
, \quad
\Gamma^{-+} (x) \equiv \sum_{w \in 2{\rm Z}+1, n \in 2{\rm Z}}
 q^{{{1}\over{2}} ({{n}\over{x}} + {{wx}\over{2}} )^2 }
 {\bar q}^{{{1}\over{2}} ({{n}\over{x}} - {{wx}\over{2}} )^2 }
 \quad . \cr
\quad && \quad
\label{eq:evenodd}
\end{eqnarray}
It is evident that the functions $\Gamma^{++}$ and $\Gamma^{--}$
are invariant while $\Gamma^{+-}$ is mapped to $\Gamma^{-+}$, and
vice versa, under both modular and thermal duality
transformations.

\end{document}